\documentclass[12pt]{article}
\usepackage{amsfonts,latexsym,fullpage,amsmath,latexsym,amssymb,epsf}

\date{May 9, 2003}

\newcommand{\cB}{{\cal B}}

\newcommand{\ket}[1]{|#1\rangle}
\newcommand{\bra}[1]{\langle #1|}

\newcommand{\beq}{\begin{equation}}
\newcommand{\eeq}{\end{equation}}
\newcommand{\beqy}{\begin{eqnarray}}
\newcommand{\eeqy}{\end{eqnarray}}

\newtheorem{Definition}{Definition}

\newtheorem{Theorem}{Theorem}

\newenvironment{Proof}{{\it Proof: \,}}{$\Box$ \vspace{0.3cm}}

\newenvironment{Definition*}{{\bf Definition}}{}

\def\C{{\mathbb{C}}}

\def\R{{\mathbb{R}}}

\newcommand{\B}{{\cal B}}
\newcommand{\cH}{{\cal H}}

\newcommand{\cV}{{\cal V}}

\begin{document}

\title{``Identity check'' is QMA-complete}

\author{Dominik Janzing\thanks{e-mail: {\protect\tt
\{janzing,wocjan\}@ira.uka.de}}, Pawel Wocjan, and
Thomas Beth \\ \small Institut f{\"u}r Algorithmen und Kognitive
Systeme, Universit{\"a}t Karlsruhe,\\[-1ex] \small Am Fasanengarten 5,
D-76\,131 Karlsruhe, Germany}

\maketitle

\abstract{We define the problem ``identity check'': Given a 
classical description of
a quantum circuit, determine whether it is  almost
equivalent to the identity.
Explicitly, the task is to decide whether 
the corresponding unitary is close to a complex multiple
of the identity matrix with respect to the operator norm.
We show that this problem is QMA-complete. 

 A generalization of this problem is ``equivalence check'': Given two
descriptions of quantum circuits and a description of a common
invariant subspace,  decide whether the restrictions of the circuits
to this subspace almost coincide. We show that equivalence check is also in
QMA and hence QMA-complete.}

\section{Stating the problem ``equivalence check''}

\label{EquiSec}
So far there is only one QMA-complete problem known, namely 
the $3$-local Hamiltonian problem \cite{KitaevShen,KempeRegev,Zoo}.
Here we give another example that occurs naturally 
in the problem of constructing quantum networks from elementary gates:

Let $U$ be a quantum network acting
on $n$ qubits that consists of two-qubit gates
\[
U=U_k  \cdots U_2 U_1\,.
\]
Someone 
claims that 
the same transformation $U$  could also be implemented by
another sequence 
\[
V_l \cdots V_2 V_1\,.
\]
Assume that he did not tell us why he thinks that this sequence also 
implements
 $U$. How difficult is it to determine whether it really does?
Also the following slight modification of the  problem is natural. 
Usually we are not interested in the whole physical state space
but rather in a computational  subspace.                            
This subspace may, for instance, be defined by a quantum error correcting code
\cite{Steane}
or a decoherence free subspace \cite{ZanardiDec,ViolaDec}. 
Then it is not relevant whether the alternative network coincides with
the original one on the whole space but only on the code space.
Assume that we already know (for example by construction) 
that the alternative network
leaves the subspace invariant.
Does the alternative circuit agree with the original one when 
it is restricted
to the subspace?
This is obviously equivalent to the question whether 
the restriction of 
\[
V_1^\dagger V_2^\dagger \cdots V_l^\dagger U_k \cdots U_2 U_1
\]
is the identity.

First we introduce some
notations that will be used trough the paper. We denote the Hilbert
space of a qubit by $\B:=\C^2$. Let $x\in\{0,1\}^*$ be an arbitrary
binary string. We denote the length of $x$ by $|x|$. For any Hilbert
space $\cH$ we denote the set of density matrices acting on $\cH$
by $S(\cH)$.

We
define formally:

\begin{Definition}[Equivalence Check]\label{EquiDef}${}$\\
Let $x,y$ be classical descriptions of  quantum networks consisting
of $poly(|x|)$ and $poly(|y|)$ many two-qubit gates, respectively.
Let $U_x$ and $U_y$ be the unitary transformations implemented by 
the circuits acting on $n$ qubits with $n=poly(|x|)$ and $n=poly(|y|)$. 
Given a common invariant subspace $\cV$  of $\cB^{\otimes n}$.
Let $\cV$ be specified by a quantum circuit $V$ on $\cB^{\otimes (n+m)}$ 
with polynomial complexity such that
$V \cV  = W_1$ where $W_1$ is  the space of all states
of $\cB^{\otimes (n+m)}$ where the last 
qubit is in the state $|1\rangle$.

The problem equivalence check is to decide whether
the restrictions of $U_x$ and $U_y$ to $\cV$ 
coincide approximatively. Explicitly we assume that it is promised that
either

\begin{enumerate}
\item There is a vector $|\Psi\rangle \in \cV$ such that
\[
\|(U_xU^\dagger_y-e^{i\phi} {\bf 1}) |\Psi\rangle \| \geq \delta
\]
for all $\phi \in [0,2\pi)$ or
 
\item
There exists an angle $\phi \in [0,2\pi)$ such that
for all vectors $|\Psi\rangle \in \cV$ 
\[
\|(U_xU^\dagger_y-e^{i\phi} {\bf 1} )|\Psi\rangle \| \leq \mu\,,
\]
\end{enumerate}
where $\delta -\mu \geq 1/poly(|x|)$ and $\delta -\mu  \geq
1/poly(|y|)$.
\end{Definition}
 
In the following section we will show that equivalence check is in QMA.
In Section~\ref{IdencompleteSec} we will show that a specific instance of
equivalence check, namely to decide whether a circuit is almost 
equivalent to the identity, encompasses QMA. Hence equivalence check
and identity check are both QMA-complete.

\section{Equivalence check is in QMA}
 
\label{EquiQMASec}

The complexity class QMA consists of the problems of deciding whether a
given string is in a certain language in QMA. The set of QMA languages
is defined following \cite{KempeRegev}.

\begin{Definition}[QMA]\label{QMA}${}$\\
Fix $\epsilon=\epsilon(|x|)$ such that $2^{-\Omega(|x|)} \leq \epsilon
\leq 1/3$.  Then a language $L$ is in QMA if for every classical input
$x \in \{0,1\}^*$ one can efficiently generate (by classical
precomputation) a quantum circuit $U_x$ (``verifier'') consisting of
at most $p(|x|)$ elementary gates for an appropriate polynomial $p$
such that $U_x$ acts on the Hilbert space
\[
\cH:= \B^{\otimes {n_x}} \otimes \B^{\otimes m_x}\,,
\]
where $n_x$ and $m_x$ grow at most polynomially in $|x|$. The first
part is the input register and the second is the ancilla register.
Furthermore $U_x$ has the property that
\begin{enumerate}
\item If $x \in L$
there exists a quantum state $\rho$ that is accepted by the
circuit with high probability, i.e.,
\[\exists \rho \in S(\B^{n_x})\,,\quad
tr(U_x\,(\rho\otimes\ket{0\ldots 0}\bra{0\ldots 0})\,U^\dagger_x\, P_1)
\geq 1-\epsilon\,,
\]
where $P_1$ is the
projection corresponding to the measurement ``Is the first qubit in
state $1$?''.
\item If $x\not\in L$ all
quantum states are rejected with high probability, i.e.,
\[
\forall\rho \in S(\B^{n_x})\,,\quad
tr(U_x\,(\rho\otimes\ket{0\ldots 0}\bra{0\ldots 0})\,U_x^\dagger\, P_1)
\leq \epsilon\,.
\]
\end{enumerate}
\end{Definition}
Note that our ``witnesses'' are mixed states in contrast to the
definitions in \cite{KitaevShen,KempeRegev}.  Due to linearity
arguments this modification does not change the language $L$.  Note
furthermore that it is always possible to construct a verifier for the
same language with $\epsilon'$ arbitrarily close to $0$. This
``amplification of probabilities'' is described in \cite{KitaevShen}
in full detail. This may be necessary in Section \ref{IdencompleteSec}.

To prove that equivalence check is in QMA we have to describe how
to give a witness state that proves that $U_x$ and $U_y$ do not
coincide.
For an arbitrary unitary operator $W$ the difference from 
multiples of the identity is a normal operator. Hence its operator norm
is given by the greatest modulus of the eigenvalues.
Therefore the 
operator norm distance between $W$ and the set of trivial
transformations (global phases) can be determined 
as follows.

Whenever there exist eigenvalues $\exp(i\alpha)$ and $\exp(i\beta)$
of $W$ the norm distance to $\exp(i\phi) {\bf 1}$ is
at least 
\begin{equation}\label{specNorm}
\max \{ |e^{i\alpha}- e^{i\phi}|, |e^{i\beta} -e^{i\phi}|\}
\end{equation}
If $|\alpha-\beta| \leq \pi$ the minimum of expression (\ref{specNorm})
is achieved for $\phi:=(\alpha-\beta)/2$ and the norm distance
to the trivial transformations implementing global phases 
is hence at least
\[
|1-e^{i(\alpha-\beta)/2}|=\sqrt{2(1-\cos((\alpha-\beta)/2))}\,.
\]
Let $U'_x,U'_y$ be the restrictions of $U_x$ and $U_y$  
to $\cV$.
If case 1  of Definition \ref{EquiDef} is true there exists
eigenvectors $|\psi_a\rangle$ and $|\psi_b\rangle$ 
of $U'_x(U'_y)^\dagger$  with eigenvalues
$e^{i\alpha}$ and $e^{i\beta}$, respectively
such that
\[
\delta
\leq \sqrt{2(1-\cos((\alpha-\beta)/2))}
\]

In order to check that the eigenvalues corresponding to the 
given eigenvectors satisfy this criterion one can use
the phase estimation procedure \cite{ClevePhase}. 

Due to the promise that in case 
2 one has $\sqrt{2(1-\cos((\alpha-\beta)/2)}\leq \mu$ 
the accuracy of the phase estimation has to be chosen such that
$\cos((\alpha-\beta)/2)$ can be determined up to 
an error of $(\delta^2-\mu^2)/4$.
It remains to check whether $|\psi_a\rangle$ and $|\psi_b\rangle$ 
are elements of $\cV$. This can be done using the given circuit
$V$.  

Actually the setting of QMA problems (see Definition \ref{QMA}) requires
that the witness is one quantum state instead of two. 
Formulated as an Arthur-Merlin game \cite{KitaevShen} Merlin proves
Arthur that a string $x$ is in QMA by sending the witness quantum state.
Here he may prove that $U_xU_y^\dagger$ has eigenvalues of non-negligible
distance by sending the state
$|\psi_a\rangle \otimes|\psi_b\rangle$.
A priori it is not clear that Merlin cannot cheat by sending entangled
(wrong) witnesses. However, one can check easily that the circuit in Fig.1 
treats any state
\[
\sum_j c_j|\psi^j_a\rangle \otimes |\psi^j_b\rangle
\]
as an incoherent mixture of product states
$|\psi^j_a\rangle \otimes |\psi^j_b\rangle$ with weights $|c_j|^2$.
Note that it is also irrelevant whether the witness states
$|\psi_a\rangle$ and $|\psi_b\rangle$ are really eigenstates of
$U_x U_y^\dagger$. The phase estimation procedure can only
produce output that really exists as eigenvalues (up to the accuracy
that is determined by the size of the used ancilla register).
In Fig.~1 one can see the whole circuit.

\begin{figure}
\centerline{
\epsfbox[0 0 455 283]{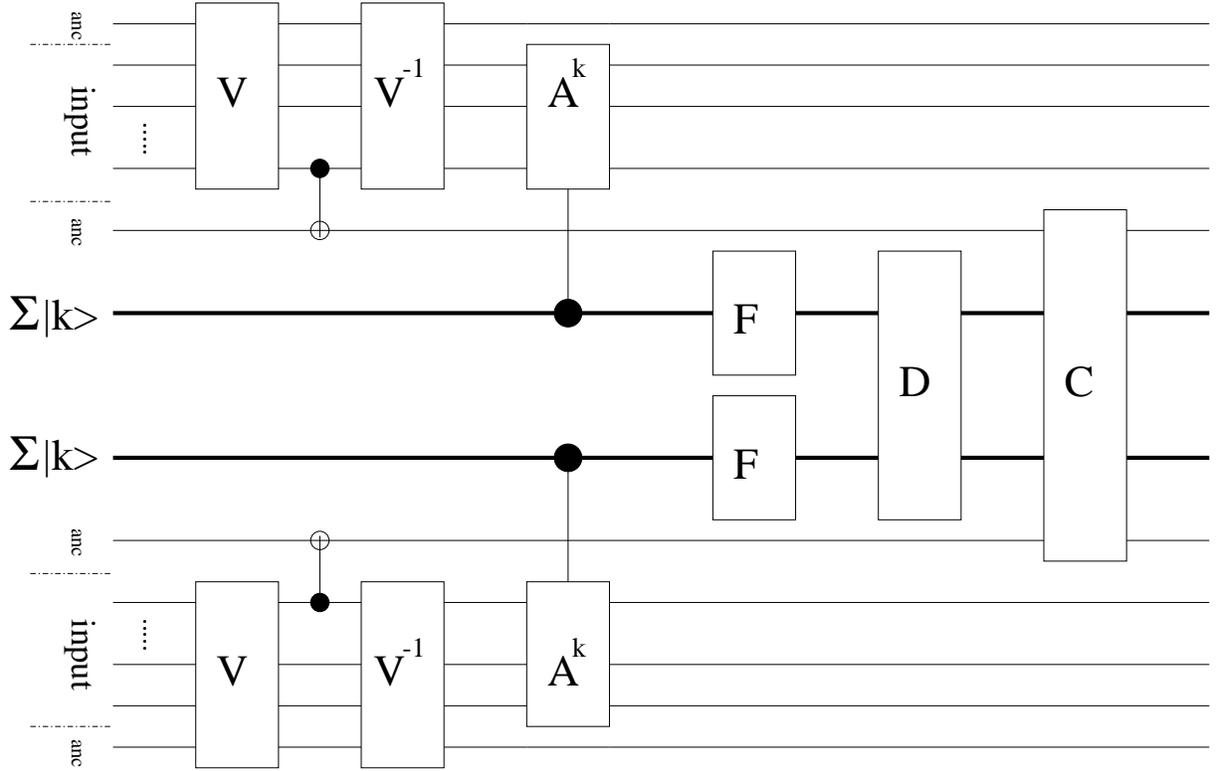}}
\caption{{\small Circuit used to verify that $U_xU_y^\dagger$ is not close
to the identity on the subspace $\cV$. 
The two copies of $V$ check
that the witness states are really elements of $\cV$. The 
results of this check are copied to  additional ancilla qubits by 
Controlled-NOT gates. 
The main part of the circuit ($A^k$ and $F$) is a usual phase
estimation procedure. 
The ancilla registers are initialized into the superposition 
 state
$(1/\sqrt{m})\sum_{k\leq m} |k\rangle$ and 
 control the implementation of
$A^k:=(U_xU_y^\dagger)^k$. The state $|k\rangle$  obtains a 
phase according to the eigenvalues of 
$A^k$. By Fourier transformations  $F$ 
the phases can be read out 
from the ancilla registers. 
A circuit $D$ computes the phase difference and $C$ checks whether
the difference is sufficiently large and 
the witness states are elements of the subspace $\cV$.}}
\end{figure}

\section{``Identity check'' is QMA-complete}

\label{IdencompleteSec}

First we state the problem ``Identity check'' formally.

\begin{Definition}[Identity Check]\label{IdenDef}${}$\\
Let $x$ be a classical description of a quantum circuit
$U_x$ of complexity polynomial in $|x|$. 
Decide whether $U_x$ is close to the trivial transformation
in the following sense.
Decide which of the two following cases is true
given the promise that either of 1. or 2. is satisfied:

\begin{enumerate}
\item for all $\phi \in [0,2\pi)$
\[
\|U_x- e^{i\phi}{\bf 1}\| \geq \delta 
\]
or 
\item
there exists an angle $\phi\in [0,2\pi)$ such that 
\[
\|U_x- e^{i\phi}{\bf 1}\| \leq \mu\,.
\]
\end{enumerate}
Assume furthermore that $\delta - \mu \geq 1/poly(|x|)$.
\end{Definition}
Note that this problem 
is a specific instance of equivalence check. 
 
The general QMA setting is that a quantum  circuit $U$ is given 
and the problem is to decide whether there is a state $|\psi\rangle$ such that
the state
\[
U |\psi \rangle \otimes |0\dots 0\rangle
\]
has the property that the first qubit is with high probability in the 
state $|1\rangle$. 
In order to show that Identity Check encompasses QMA 
we construct a circuit $Z$ that implements a unitary close to the identity 
whenever there is no state that is accepted by $U$ and a circuit 
less close to the identity if there is a witness.
The register is extended by one qubit and the whole circuit is the 
transformation 
\[   
Z:=U^\dagger W  U  V\,.
\]
The transformation $V$ is a  phase shift 
controlled by the states of the ancillas. Whenever the ancilla part of
 the register is initialized in the state $|0\dots0\rangle$ 
the additional qubit gets a phase $\exp(i\varphi)$.
The gate $W$ is a phase shift controlled by the output qubit of $U$.
The additional qubit gets a phase $\exp(i\varphi)$  whenever
the circuit has accepted (see Fig.2).

\begin{Theorem}
Let $U$ be a quantum circuit on $\cB^{\otimes (n+m)}$ 
with the promise that either of two cases in Definition \ref{QMA} is
true. 
Then for the circuit $Z$ in Fig.~2 the following statements hold:

\begin{figure}
\centerline{
\epsfbox[0 0 312 204]{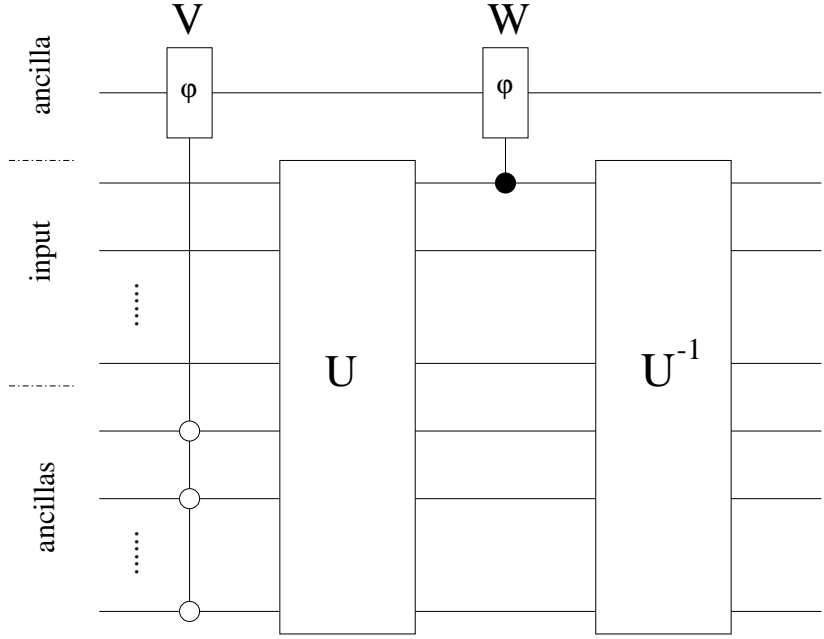}}
\caption{{\small Circuit $Z$
 consisting of $U,U^\dagger$ and two controlled phase
shifts $V$ and $W$ with phase $\varphi$. 
If $U$ rejects all states with high probability the circuit
is closer to the identity than in the case that there is a state
that is likely to be accepted. The first ancilla can only obtain a phase
shift $2\varphi$ if the other ancilla register has been  correctly initialized
and the input has been accepted by $U$.}}
\end{figure}

If case 1 is true then we have 
\[
\|Z-e^{i\gamma} {\bf 1}\| \geq \sqrt{2(1-\cos\varphi)} -2 \sqrt{\epsilon}
\]
for all $\gamma \in \R$. 

If case 2 is true then we have
\[
\|Z- e^{i\varphi/2}{\bf 1}\|\leq 2\sqrt{1-\cos(\varphi/2)} +2 \sqrt{2\epsilon}
\]
\end{Theorem}

\begin{Proof}
The effect of $Z$  
 on a general state $|\Psi\rangle$ can
be understood if we express $|\Psi\rangle$ as
\[
|\Psi\rangle=|\Psi_1\rangle \oplus |\Psi_2\rangle\,,
\]
where 
$|\Psi_1\rangle$ is a state with ancillas all set to $0$ and
 $|\Psi_2\rangle$ a state with ancilla register in states different from
$|0\dots 0\rangle$.
We have
\[
Z|\Psi\rangle =  U^\dagger W  U  V|\Psi_1\rangle \oplus U^\dagger W  U  V|\Psi_2\rangle\,.
\]
Consider case 2 and the effect of $Z$ on the summand $|\Psi_1\rangle$:
\[
U^\dagger W U V |\Psi_1\rangle =
U^\dagger W P_1 U V |\Psi_1 \rangle
\oplus U^\dagger W ({\bf 1}-P_1) U V |\Psi_1 \rangle
\]
where $P_1$ is (see Definition \ref{QMA}) the projection onto the state
$|1\rangle$ of the output qubit. By definition of $W$ one has
\[
W({\bf 1}-P_1)=({\bf 1}-P_1)\,.
\]
Hence we have
\[
Z|\Psi_1\rangle =U^\dagger W P_1 U V |\Psi_1 \rangle \oplus
U^\dagger ({\bf 1}-P_1) U V |\Psi_1 \rangle
= U^\dagger W P_1 U V |\Psi_1 \rangle +
 V |\Psi_1 \rangle - U^\dagger P_1 U V |\Psi_1\rangle
\]
Since the probability of acceptance is at most $\epsilon$ the length of
the vector $P_1 U V |\Psi_1 \rangle$ is at most 
$\sqrt{\epsilon}\|\,|\Psi_1\rangle\|$.
We conclude 
\begin{equation*}
\|Z|\Psi_1\rangle -V|\Psi_1\rangle \|\leq 2\sqrt{\epsilon}\|\,|\Psi_1\rangle\|\,.
\end{equation*}

Note that $\|V-\exp(i\varphi/2){\bf 1}\|=|1-\exp(i\varphi/2)\|$
due to the arguments at the end of Section~\ref{EquiQMASec}.
Due to $\|V|\Psi_1\rangle -e^{i\varphi/2}|\Psi_1\rangle\| 
\leq |1-\exp(i\varphi/2)| \,\|\,|\Psi_1\rangle \|$ we
have
\begin{equation}\label{psi1}
\|Z|\Psi_1\rangle - e^{i\varphi/2}|\Psi_1\rangle\| \leq (2\sqrt{\epsilon}
+|1-\exp(i\varphi/2)|)\,\|\,|\Psi_1\rangle\|\,.
\end{equation}
Consider the effect of $Z$ on $|\Psi_2\rangle$.
\begin{eqnarray*}
\|Z|\Psi_2\rangle -e^{i\varphi/2}|\Psi_2\rangle 
\|&=&\|U^\dagger W U V|\Psi_2\rangle -e^{i\varphi/2}|\Psi_2\rangle\|\\
&=&\|U^\dagger (W- e^{i\varphi/2}{\bf 1}) U |\Psi_2\rangle\| \leq
\|W-e^{i\varphi}{\bf 1}\|\, \||\Psi_2\rangle\|\,.
\end{eqnarray*}
Together with inequality (\ref{psi1})
we have
\[
\|Z|\Psi\rangle -e^{i\varphi/2}|\Psi\rangle\| \leq
(|1-\exp(i\varphi/2)| +2\sqrt{\epsilon}) (\|\,|\Psi_1\rangle\|+|\Psi_2\rangle \|)
\leq \sqrt{2}(|1-\exp(i\varphi/2)| +2\sqrt{\epsilon})\,.
\]
With $|1-\exp(i\varphi/2)|=\sqrt{2(1-\cos\varphi/2)}$ 
we have 
\[
\|Z- e^{i\varphi/2}{\bf 1}\| \leq 2\sqrt{1-\cos(\varphi/2)} +2 \sqrt{2\epsilon} \,.
\]

Consider case 1. Let $|\psi\rangle$ be a state that is accepted by $U$ with
probability $1-\epsilon$.
Define $P_0:={\bf 1}-P_1$.
We take the state vector
\[
|\Psi\rangle:= \frac{1}{\sqrt{2}} (|0\rangle +
|1\rangle) \otimes |\psi\rangle \otimes |0\dots 0\rangle \,.
\] 
We have
\begin{eqnarray*}
Z|\Psi\rangle&=& U^\dagger W U V \frac{1}{\sqrt{2}} (|0\rangle +
|1\rangle) \otimes |\psi\rangle \otimes |0\dots 0\rangle\\&=&
U^\dagger W U \frac{1}{\sqrt{2}} (|0\rangle + e^{i\varphi}
|1\rangle) \otimes |\psi\rangle \otimes |0\dots 0\rangle \\
&=&U^\dagger W ({\bf 1} -P_0) U \frac{1}{\sqrt{2}} (|0\rangle + e^{i\varphi}
|1\rangle) \otimes |\psi\rangle \otimes |0\dots 0\rangle 
+ \\&& U^\dagger W P_0 U \frac{1}{\sqrt{2}} (|0\rangle + e^{i\varphi}
|1\rangle) \otimes |\psi\rangle \otimes |0\dots 0\rangle \\
&=&U^\dagger ({\bf 1} -P_0) U \frac{1}{\sqrt{2}} (|0\rangle + e^{i2\varphi}
|1\rangle) \otimes |\psi\rangle \otimes |0\dots 0\rangle 
+\\&& U^\dagger W P_0 U \frac{1}{\sqrt{2}} (|0\rangle + e^{i\varphi}
|1\rangle) \otimes |\psi\rangle \otimes |0\dots 0\rangle\\ &=&
 \frac{1}{\sqrt{2}} (|0\rangle + e^{i2\varphi}
|1\rangle) \otimes |\psi\rangle \otimes |0\dots 0\rangle 
-\\&& U^\dagger P_0 U \frac{1}{\sqrt{2}} (|0\rangle + e^{i2\varphi}
|1\rangle) \otimes |\psi\rangle \otimes |0\dots 0\rangle 
+\\&& U^\dagger P_0 U \frac{1}{\sqrt{2}} (|0\rangle + e^{i\varphi}
|1\rangle) \otimes |\psi\rangle \otimes |0\dots 0\rangle\\
&=:& |\hat{\Psi}\rangle -|\varphi_1\rangle +|\varphi_2\rangle\,.
\end{eqnarray*}
Note that the vectors $|\varphi_1\rangle$ and $|\varphi_2\rangle$ 
have at most norm $\sqrt{\epsilon}$ due to the high probability of
acceptance.  One checks easily that
\[
\min_{\gamma \in \R}
\|\,|\hat{\Psi}\rangle - e^{i\gamma}|\Psi\rangle \| =
\|\,|\hat{\Psi}\rangle - e^{i\varphi}|\Psi\rangle \| =
|1-\exp(i\varphi)|\,.
\]
We conclude
\[
\min_{\gamma \in \R}\|Z|\Psi\rangle - e^{i\gamma}|\Psi \rangle \|
\geq |1-\exp(i\varphi)| -2\sqrt{\epsilon}\,.
\]
With $|1-\exp(i\varphi)|=\sqrt{2(1-\cos\varphi)}$ 
we conclude that
the minimal norm difference between $Z$ and $e^{i\gamma} {\bf 1}$ is
at least
\[
\sqrt{2(1-\cos\varphi)} -2\sqrt{\epsilon}\,.
\]
\end{Proof}

As mentioned in the remark after Definition \ref{QMA}
$\epsilon$ can be made arbitrarily small. 
For small $\varphi$ the lower and upper bounds on the norm distances
between $U$ and the trivial transformations are approximatively given by
\[
\varphi+ 2\sqrt{2\epsilon}
\]
and   
\[
\sqrt{2} \varphi -2\sqrt{\epsilon}\,,
\]
respectively. This shows that for sufficiently 
small $\epsilon$ there is a sufficient separation between the lower and upper 
bound. This shows that every oracle that is able to decide
whether $Z=U_x^\dagger W U_x V$ is close to a trivial transformation
can be used to decide whether $x$ is in $L$.

\section*{Acknowledgments}

Thanks to Thomas Decker for helpful discussions.
This work was supported by grants of the BMBF-project  01/BB01B.

\end{document}